\documentclass[12pt]{article}
\usepackage{amssymb,amsmath,bm}
\usepackage{epsf}
\usepackage{epsfig,cite}
\usepackage{color}
\allowdisplaybreaks[2]

\begin{document}
\begin{titlepage}
\vspace*{-0.5truecm}


\begin{flushright}
TUM-HEP-703/08
\end{flushright}

\vspace{1truecm}

\begin{center}
\boldmath

{\LARGE\textbf{Soft-Collinear Effective Theory: \\[0.3em]
Recent Results and Applications}}

\unboldmath
\end{center}

\vspace{0.4truecm}

\begin{center}
{\bf Thorsten Feldmann}
\vspace{0.4truecm}

{\small
 {\sl Physik Department, Technische Universit\"at M\"unchen,
James-Franck-Stra{\ss}e 2, \\D-85748 Garching, Germany}\vspace{0.2truecm}
}

\end{center}

\vspace{0.6cm}
\begin{abstract}
\vspace{0.2cm}\noindent
Soft-collinear effective theory (SCET) has become a
 standard tool to study the factorization of short- and long-distance effects in processes involving low-energetic (soft) particles and high-energetic/low-virtuality
(collinear) modes. In this contribution I give a brief overview on recent results for inclusive and exclusive $B$ decays and on applications in collider physics.
\end{abstract}

\vfill 
\begin{flushright}
{ \footnotesize
[Contributed to ``Quark Confinement and the Hadron Spectrum'',
Sep 2008, Mainz, Germany]}
\end{flushright}
\end{titlepage}

\setcounter{page}{1}
\pagenumbering{arabic}


\section{Factorization and SCET}

Our ability to provide precise theoretical predictions for
high-energy processes in particle physics heavily relies on the
concept of factorization, i.e.\ the systematic separation of 
dynamical effects from short and long distances. Especially
for strong interactions -- if factorization holds --
the effects of heavy particles and/or highly virtual radiative 
corrections can be calculated in perturbative Quantum Chromodynamics (QCD), 
while the long-distance physics of light quarks and gluons can be encoded in (process-independent) hadronic matrix element of composite operators, which
can be further studied using non-perturbative methods.
A general feature of factorization is the appearance of a factorization
scale $\mu$ that relates the infrared (IR) divergences, appearing in loop
corrections to short-distance amplitudes/cross sections,
and the ultraviolet (UV) divergences
of composite operators defining the long-distance matrix elements, such that
the scale dependence cancels to any given order in perturbation theory.


A particularly interesting situation arises in processes like, for instance,
$B \to X_s \gamma$, where $X_s$ denotes a hadronic jet containing
a light strange quark with energy of order $m_b/2$ and invariant mass of
order $\sqrt{\Lambda_{\rm QCD} m_b}$. Here, the infrared divergences of the short-distance $b \to s \gamma$ vertex corrections
can be identified as coming from quarks and gluons being either 
soft ($|k^\mu| \sim \Lambda_{\rm QCD}$) 
or collinear to the hadronic jet ($k^\mu \parallel p_X^\mu$).
The interactions of the $b$-quark with soft degrees of freedom can be
expanded in the small parameter $\Lambda_{\rm QCD}/m_b$,
and the remaining non-analytic dependence on the $b$-quark mass $m_b$
can be calculated within the well-known heavy-quark effective theory (HQET). 
The presence of additional collinear modes leads to new phenomena \cite{Neubert:1993ch}:
\begin{itemize}
 \item The $b \to s$ form factors contain Sudakov double 
       logarithms $\ln^2(p_X^2/m_b^2)$.
 \item The propagation of a collinear quark in the soft background is described by 
       a jet function.
 \item The partial rate depends on the residual 
       momentum of the $b$-quark, which is encoded in a so-called
       shape function (SF), i.e.\ the parton distribution function 
       (PDF) for the $B$-meson.
\end{itemize}
Again, the expansion in $1/m_b$ 
can be formalized in terms of an effective theory, 
SCET \cite{Bauer:2000yr,Bauer:2008}.
To this end, one includes \emph{separate} field operators for soft and collinear
modes, with soft-collinear vertices being 
multi-pole expanded according to the power-counting of momenta/wave-lengths
in the different light-cone directions \cite{Beneke:2002ni}. 
The short-distance coefficient
functions and the jet function can be calculated by perturbative matching calculations.
The renormalization-group (RG) running in SCET resums the large Sudakov
logarithms between the hard scale ($m_b$) and the jet scale ($|p_X|$)
\cite{Bauer:2000ew},
where one finally matches onto (non-local) HQET operators that define
the $b$-quark PDF.

\section{SCET applications}

While SCET originally has been designed to discuss factorization
in inclusive and exclusive $B$-decays, it has also led to some
new insights in collider physics applications. This includes,
the traditional field of QCD jet physics and parton showers 
(which is discussed in more detail by Christian Bauer in these proceedings), 
as well as resummation effects in high-energy electroweak processes.
In the following, I will present a personal selection of recent results,
illustrating the main SCET activities for $B$-decays and collider physics
(a good overview can also be obtained from the talks presented at the
recent SCET workshop 2008 \cite{SCET}).

\subsection{Inclusive $B$ decays}

\label{sec:incl}

Factorization theorems 
(see e.g.\ \cite{Korchemsky:1994jb,Akhoury:1995fp,Bauer:2003pi,Bosch:2004th})
play a key role in the determination
of the CKM matrix element $|V_{ub}|$ from inclusive semi-leptonic
$B \to X_u\ell\nu$ decays, as well as for tests of the 
Standard Model in rare penguin decays $B \to X_s\gamma$. 
In the former, one becomes sensitive to the
SF when applying the constraint
$E_X - |\vec p_X| \leq \Delta < M_D^2/M_B \,,$
in order to suppress the background from $b\to c\ell\nu$ decays.
In the latter, one is experimentally restricted to sufficiently large
photon energies, which again implies large recoil energy to the
hadronic jet.\footnote{Cuts on the jet mass $M_X$ also induce SF-sensitivity in
$B \to X_s \ell^+\ell^-$ \cite{Lee:2005pwa}.} 
In both cases, one ends up with a factorization 
theorem for the decay spectrum, which schematically reads
\begin{equation}
 d\Gamma \sim H \cdot J \otimes S \,.
\end{equation}
Here $H$ denotes the hard function, obtained from a QCD
matching calculation, which is presently known to NNLO accuracy
both, for $b \to u\ell\nu$ \cite{Bonciani:2008wf}
and $b\to s\gamma$ \cite{Melnikov:2005bx}.
Furthermore, $J$ represents the universal jet function in SCET, whose
NNLO expression (for massless quarks) has been derived in \cite{Becher:2006qw}
(for massive quarks, the NLO jet function has been given in
\cite{Boos:2005qx}, see also \cite{Fleming:2007xt}). It is convololuted
with a soft function, $S$, which denotes the $b$-quark 
SF in HQET, whose 2-loop evolution has been studied 
in \cite{Becher:2005pd}. Sub-leading SFs, entering at the level
of $1/m_b$ corrections, have been classified in \cite{Lee:2004ja}. 
We should also mention that SF-independent relations between $B \to X_s\gamma$
and $B \to X_u\ell\nu$ can be obtained by appropriately
re-weighting the experimental decay spectra, with weight functions determined
from the perturbative short-distance functions in the factorization theorem
\cite{Lange:2005qn}.

\subsubsection{The $B$-meson shape function}

For the community of this workshop, the perhaps most interesting ingredient
is the $B$-meson SF, which is defined via the light-cone matrix element
(with HQET fields $h_v$)
\begin{equation}
 \widehat S(\hat \omega=\bar\Lambda-\omega) = \langle B | \bar h_v \, \delta(\omega-i n \cdot D) \, h_v |B\rangle \,,\qquad \mbox{($n^2=0,\ n\cdot v=1,\ \bar\Lambda=m_B-m_b$)} \,.
\end{equation}
The SF has support for $0 \leq \hat \omega \leq \infty$, where large values of 
the (residual) light-cone momentum $\hat \omega$ are described by a 
radiative tail which can be calculated in perturbation theory.

\begin{figure}[t!!!]
\centerline{(a) \hspace{-1em} \includegraphics[height=0.19\textheight]{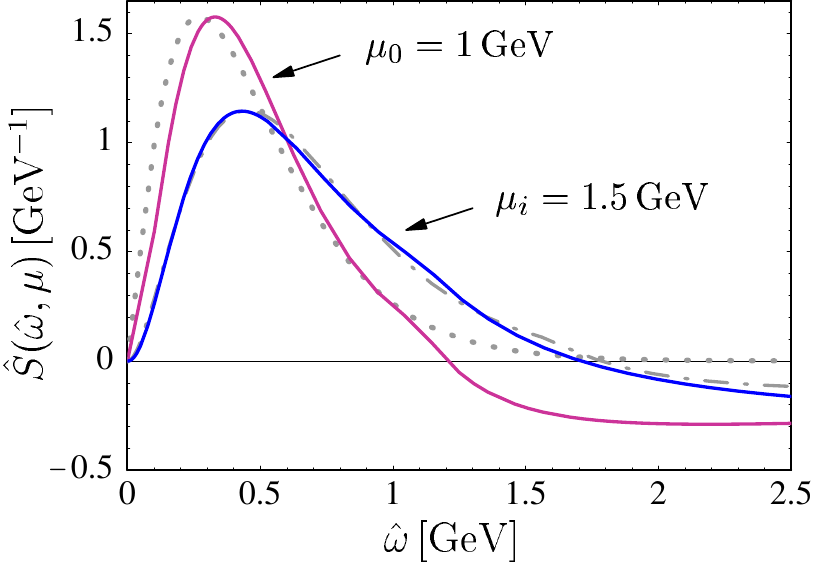}
\quad (b) \hspace{-1em}  \includegraphics[height=0.19\textheight]{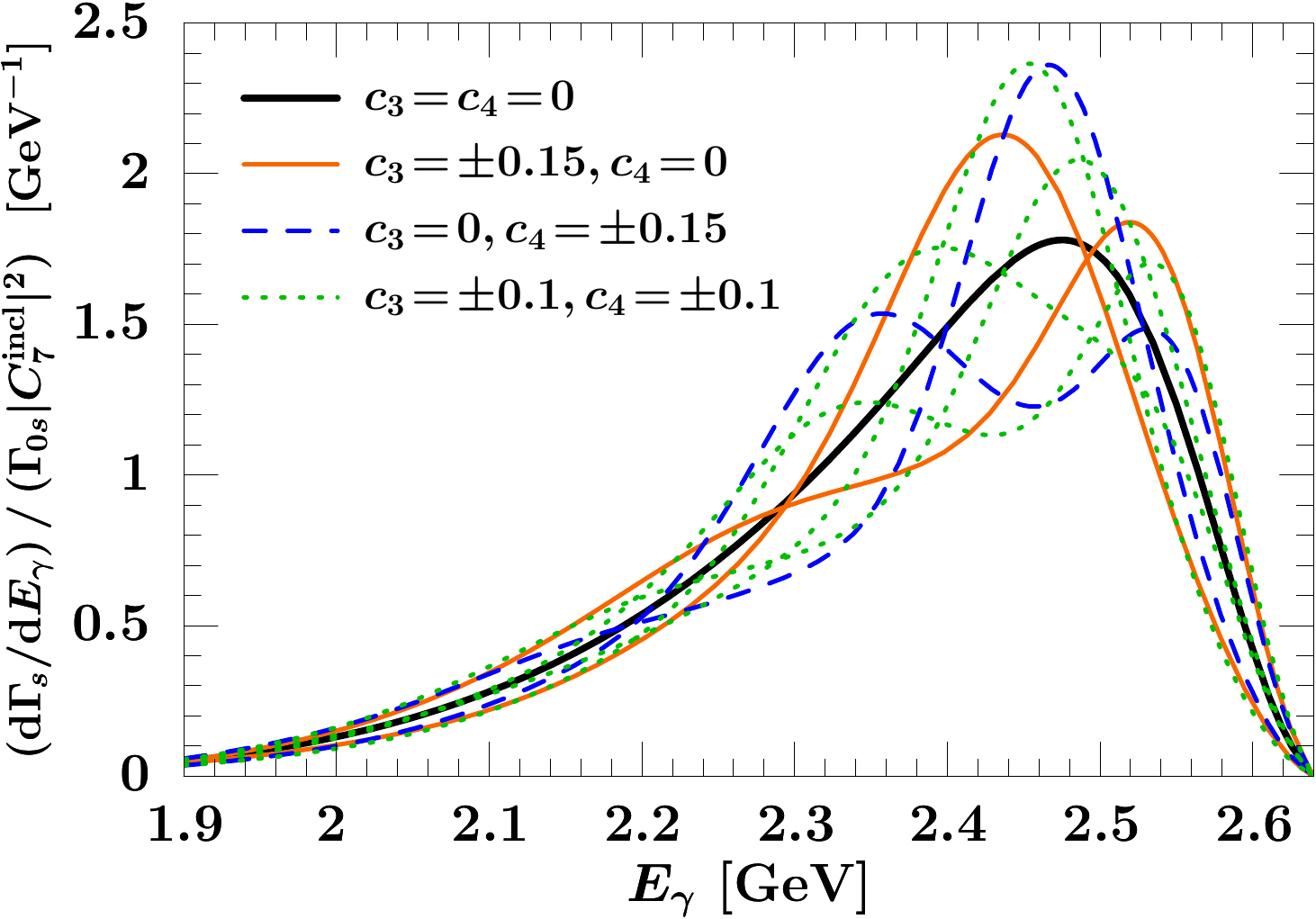}}
\caption{(a) Model SF at low input scale $\mu_0=1$~GeV and after evolution 
(solid lines, figure taken from \cite{Bosch:2004th}). For the meaning of 
the dotted curves and further details, see section 9.2 in \cite{Bosch:2004th}.
(b) Photon spectrum in $B\to X_s\gamma$ resulting from different
   profile functions in (\protect\ref{alt}). Figure taken from \cite{Ligeti:2008ac}.
\label{fig:SF1}}
\end{figure}

Experimentally, the SF can be directly constrained by the measured photon spectrum
in $B \to X_s\gamma$ decays (via the above factorization theorem).
In addition the moments of the $B \to X_c\ell\nu$ spectra determine
the HQET parameters $\bar\Lambda, \mu_\pi^2,\ldots$, which in turn constrain
the moments of the SF in a given factorization scheme.
For instance, the authors of \cite{Bosch:2004th} 
propose a model-parametrization for the SF at low input scales 
(in the so-called SF scheme),
\begin{equation}
 \widehat S(\hat \omega,\mu_0) = 
\frac{N}{\Lambda} \left(\frac{\hat\omega}{\Lambda}\right)^{b-1} \exp\left(-b \, \frac{\hat \omega}{\Lambda} \right) + \frac{\alpha_s}{\pi} \times \mbox{[radiative tail]} \,,
\end{equation}
where $N$ is a normalization factor, 
and the free parameters ($b,\Lambda$) can be related to 
$\bar\Lambda$ and $\mu_\pi^2$.  An example is plotted in Fig.~\ref{fig:SF1}(a).
The ansatz can be compared to the $B \to X_s\gamma$ spectrum predicted by
the factorization formula.

An alternative approach has recently been proposed in \cite{Ligeti:2008ac}.
One starts with the perturbative result for the \emph{partonic} SF,
 $\widehat S_{\rm part.}(\hat \omega,\mu_0)=\delta(\hat\omega) 
+ \frac{\alpha_s}{\pi} \, [\cdots]$, and generates model SFs via
\begin{equation}
 \widehat S(\hat \omega,\mu_0) := \int dk \, \widehat S_{\rm part.}(\hat\omega-k,\mu_0)
 \, \widehat F(k) \,.
\label{alt}
\end{equation}
The profile function $\widehat F(k)$ can be directly normalized
to HQET parameters and expanded in terms of suitable basis functions;
examples are shown in Fig.~\ref{fig:SF1}.
This procedure is expected to be advantageous for systematic studies
of theoretical uncertainties in global fits to $B \to X_s\gamma$ spectra
and $B \to X\ell\nu$ moments.

\subsubsection{Theoretical limitations in $B\to X_s\gamma$}

In the theoretical discussion of the $B \to X_s\gamma$ spectrum,
a particular complication arises due to the fact that the weak effective
Hamiltonian contains operators that contribute in a different
way to the hadronization process, namely 
the chromomagnetic operator ${\cal O}_8(b \to sg)$,
and the 4-quark operators ${\cal O}_{1-6}(b \to s q\bar q)$. At sub-leading
order in the $1/m_b$ expansion this leads to qualitatively new effects, where
the photon does not couple directly to the short-distance $b \to s$ transition.
This requires a new type of factorization theorem, which involves a new jet function in the
direction opposite to $p_X$, as well as new soft functions from operators
that are non-local with respect to two light-cone directions
\cite{GilPaz}.
On the one hand, these effects are difficult to estimate (the
vacuum insertion approximation leads to corrections of order 5\%).
On the other hand, they provide a potential mechanism for the observation
of CP violating effects, and the leading mechanism for isospin asymmetries
between decays of charged and neutral $B$-mesons in this channel.

\subsection{Exclusive $B$ decays}

\label{sec:excl}

SCET applications in exclusive $B$ decays reveal some new aspects
compared to the inclusive case. First of all, it has to be realized
that the decay into a \emph{few} light energetic hadrons (with 
mass $m^2 \sim {\cal O}(\Lambda^2)$) is
power-suppressed compared to the production of a generic jet 
(with mass $m_X^2 \sim {\cal O}(\Lambda m_b)$), since it requires
a particular fine-tuning in the phase space of the $B$-meson 
spectator system. A related subtlety arises from so-called 
endpoint divergences which prevent the complete (perturbative)
factorization of soft and collinear
modes (with small invariant mass $\sim m^2$). 
Factorization theorems for exclusive heavy-to-light amplitudes 
thus take the generic form \cite{Beneke:1999br}
\begin{equation}  
{\cal A}_i(B \to M M') =
 {\xi_{M}} \cdot
  T_i^{\rm I} \otimes {\phi_{M'}} \, 
 + T_i^{\rm II} \otimes {\phi_B} \otimes {\phi_M}
  \otimes {\phi_{M'}} + \ldots \,,
\label{fact.excl}
\end{equation}
where $M,M'$ denote light mesons in the final state.\footnote{The case
with photons and/or lepton-pairs in the final state can be described
in a similar way \cite{Korchemsky:1999qb,Beneke:2000wa,Beneke:2003pa,Beneke:2001at,Lange:2003pk}.}
Here $T_i^{\rm I,II}$ are short-distance functions, where 
the $T_i^{\rm II}$  
further factorize into a hard and an exclusive jet function 
(including spectator scattering), 
but the $T_i^{\rm I}$ do not \cite{Beneke:2003pa,Hill:2002vw}.
Furthermore, $\xi_M$ denotes a universal form factor for $B \to M$
transitions, and $\phi_{M,B}$ are
light-cone distribution amplitudes (LCDAs) for light and heavy hadrons.
Again, $1/m_b$ corrections introduce new factorizable and non-factorizable terms.
Recent perturbative calculations include NNLO corrections to
$T_i^{\rm I}$ in non-leptonic $B$ decays \cite{Bell:2007tv},
NLO spectator scattering in non-leptonic $B$ decays ($T_i^{\rm II}$)
for tree amplitudes  \cite{Beneke:2005vv}
and the leading penguin amplitudes \cite{Beneke:2006mk},
as well as ${\cal O}(\alpha_s^2)$ corrections from  
${\cal O}_7^\gamma$ and ${\cal O}_8^g$ in $B \to V\gamma$ decays
\cite{Ali:2007sj}.
Below, let us again have a closer look at the
non-perturbative ingredients related to $b$-hadrons
in the factorization formula (\ref{fact.excl}).

\subsubsection{Light-cone distribution amplitudes for $b$-hadrons}

\begin{figure}[t!!]
 \begin{center}
  \includegraphics[height=0.19\textheight]{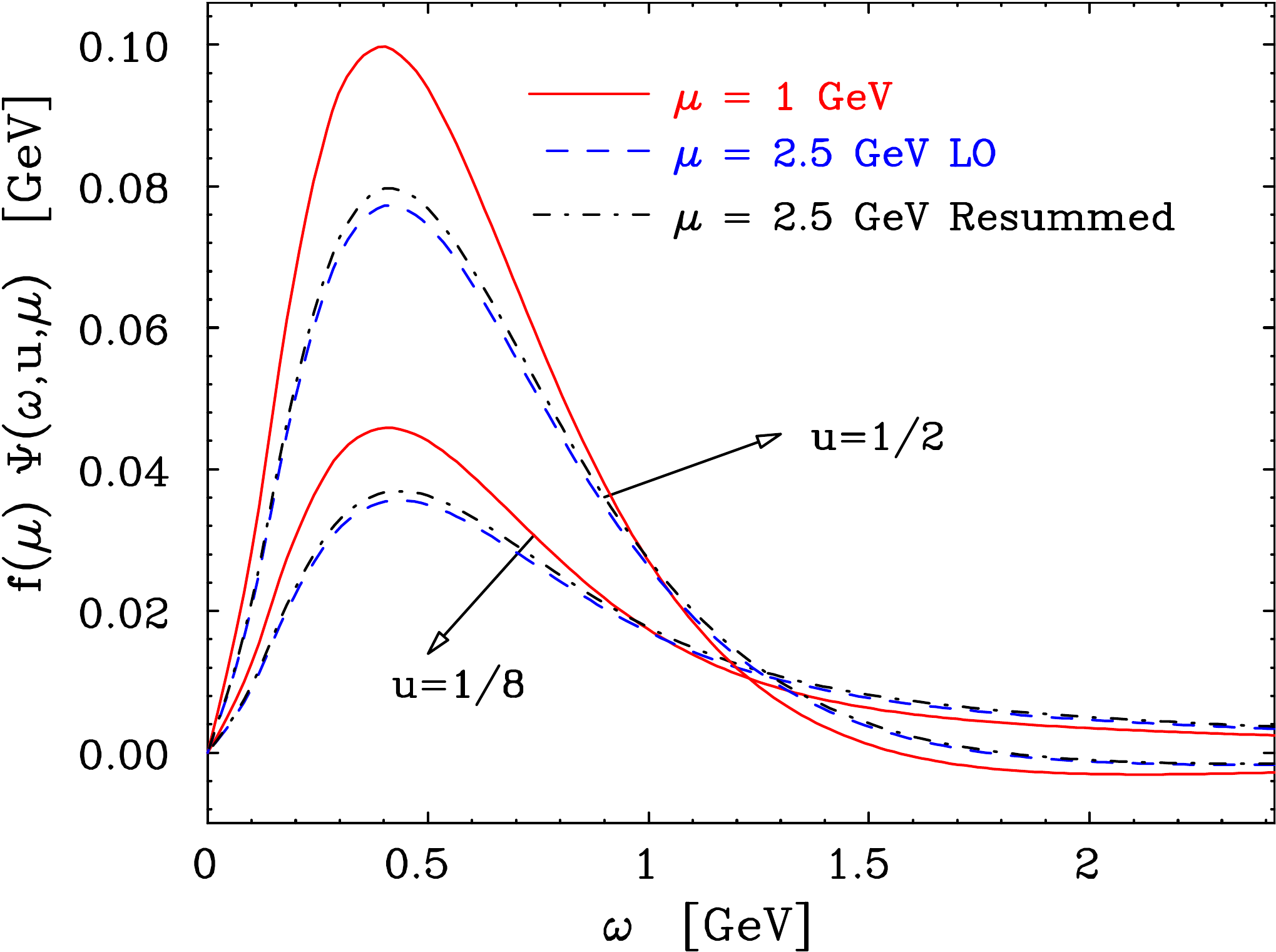} \qquad 
  \includegraphics[height=0.19\textheight]{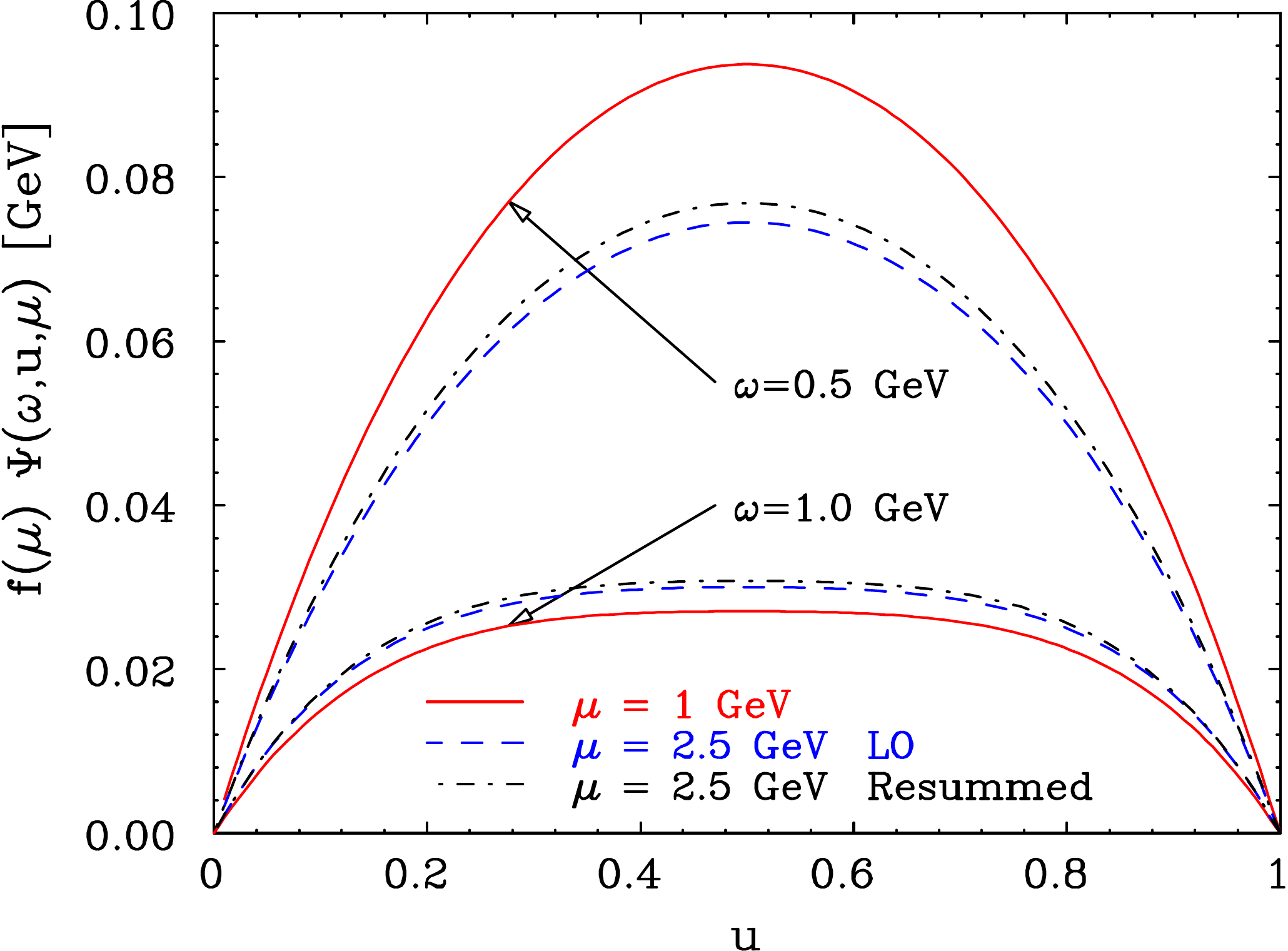}
 \end{center}
\caption{\label{fig:PhiBaryon} LCDA for $\Lambda_b$ as a function
   of $\omega=\omega_1+\omega_2$ and $u=\omega_1/\omega$ for different scales.
   Figure from \cite{Ball:2008fw}.}
\end{figure}

2-particle LCDAs for $B$-mesons are defined from non-local matrix elements in HQET \cite{Grozin:1996pq}
\begin{equation}
 \langle 0| \bar q(z)_\beta \, [z,0] \, h_v(0)_\alpha | B(v)\rangle
\qquad (z^2=0) 
\end{equation}
(see also \cite{Beneke:2000wa}) where $[z,0]$ is a gauge link.
It can be expressed in terms of two functions $\phi_B^\pm(\omega,\mu)$, where
$\omega$ represents the light-cone momentum of the spectator quark.
The 1-loop evolution kernel for $\phi_B^+$ has been derived in \cite{Lange:2003ff}. 
Of particular importance is the inverse moment $\langle \omega^{-1}\rangle_B^+$
which appears in the LO expression for $T_i^{\rm II}$ in (\ref{fact.excl}).
Its value has been estimated from QCD sum rules \cite{Braun:2003wx}, yielding
$\langle \omega^{-1} \rangle_B^+\big|_{\mu=1~{\rm GeV}} = (2.15 \pm 0.5)/{\rm GeV}$,
and from a moment analysis \cite{Lee:2005gza}, which results in
$\langle \omega^{-1} \rangle_B^+\big|_{\mu=1~{\rm GeV}} = (2.09 \pm 0.24)/{\rm GeV}$.
General properties of the $B$-meson LCDAs (evolution equations,
equations-of-motion constraints, radiative tail) can also
be verified by assuming a non-relativistic bound state at low scales,
and explicitly calculating radiative corrections from relativistic gluon exchange
\cite{Bell:2008er}.

Recently, also a systematic study of LCDAs for $\Lambda_b$ baryons 
appeared \cite{Ball:2008fw}. The 3-particle LCDAs are functions of the light-cone
momenta $\omega_{1,2}$ for the two spectator quarks. The evolution
equation for the ``leading-twist'' LCDA
contains a piece related to the Lange-Neubert kernel
\cite{Lange:2003ff} which generates a radiative tail when either
of the two momenta $\omega_{1,2}$ is large, and a piece related to
the ERBL kernel \cite{ERBL}, which redistributes the momenta within
the spectator di-quark system. Modelling the leading LCDA at low
scales with the help of sum rules, one obtains the shapes illustrated
in Fig.~\ref{fig:PhiBaryon}.

\subsubsection{The universal form factors $\xi_M$}

The universal form factors $\xi_M$ are
factorization-scale and scheme-dependent quantities.
In a simple physical factorization scheme \cite{Beneke:2000wa}, one identifies
$\xi_M$ with one of the $B \to M$ transition form factors in QCD, and uses
standard non-perturbative methods (QCD light-cone sum rules, lattice) to
estimate their size.
An alternative way is to use a definition in SCET \cite{Beneke:2003pa},
which for decays into light pseudoscalars $P$ with large recoil-energy $E$ reads
\begin{equation}
 \langle P(E)| \bar \xi W_c \, Y_s h_v |B(v)\rangle_\mu = 2 E \, \xi_P(E,\mu) \,,
\label{xiM}
\end{equation}
where $\xi$ is a collinear light-quark field in SCET,
and the Wilson lines $W_c$ and $Y_s$ appear to render the definition invariant under
independent collinear and soft gauge transformations.
A non-perturbative estimate of the so-defined form factors can be obtained from
sum-rules based on correlation functions in SCET \cite{DeFazio:2005dx}.
The correlators again factorize into a perturbative kernel and
light-cone distribution amplitudes for the $B$-meson. 
For instance, considering the correlator with an axial-vector current
to interpolate a (massless) pion, one obtains at
tree-level,
\begin{equation}
 \Pi^{(0)}(\omega',\mu)  =  f_B m_B \, \int_0^\infty d\omega \,
       \frac{\phi_B^-(\omega,\mu)}{\omega -\omega' - i \eta} \,,
\end{equation}
which provides the leading term in 
the sum rule (see also \cite{Khodjamirian:2005ea})
\begin{equation} 
 m_b \, f_\pi \, \xi_\pi(E,\mu)  = \frac{1}{\pi} \,
    \int_0^{{\omega_s}} d\omega' \, e^{-\omega'/{ \omega_M}}
   \, {\rm Im}\left[\Pi(\omega',\mu)\right] \,,
\end{equation}
where $\omega_s = s_0/2E$ is a threshold parameter characterizing the
onset of the continuum, and $\omega_M=M^2/2E$ is the Borel parameter.
As for any sum-rule calculation, the intrinsic uncertainties of the
procedure have to be estimated by a carefully defined optimization
procedure for the sum-rule parameters, together with an evaluation
of sub-leading effects from higher-order radiative and $1/m_b$-corrections.
At present, the SCET sum-rule results \cite{DeFazio:2005dx}
include ${\cal O}(\alpha_s)$ corrections,
but no power-suppressed effects, and typically have relative uncertainties of
order $25\%$, where a significant part of the error stems from the poor knowledge
of the $B$-meson decay constant and the LCDA $\phi_B^-$.

\subsection{Collider applications}

\label{sec:ew}

SCET cannot only be applied to $B$-decays, but also helps to
systematically study radiative corrections for other high-energy
processes involving soft and collinear modes. In particular, at LHC energies,
electro-weak (EW) corrections involving Sudakov logarithms have generic size 
$$
 \frac{\alpha}{4\pi \sin^2\theta_W} \, \ln^2\left[s/M_{W,Z}^2\right] \sim 15\%
\quad (@ \sqrt s\sim 4~{\rm TeV})
$$
and are thus important for precision measurements
\cite{Ciafaloni:1998xg}.
In the following
I will discuss two examples: (i) The resummation of EW
Sudakov logarithms \cite{Chiu:2007yn}, where the effective-theory approach
substantially simplifies the discussion for the spontaneously broken SM gauge
group (for applications to other high-energy
processes see also \cite{Chiu:2007dg}). (ii) Dynamical threshold enhancement
in Drell-Yan production \cite{Becher:2007ty} (see also \cite{Idilbi:2005ky}), 
where an effective soft scale appears
due to the strong fall-off of the parton distribution functions as $x \to 1$.

\subsubsection{Electroweak Sudakov logarithms}

The Sudakov form factor is defined by
the on-shell matrix element of some 2-particle operator,
$F(s=(p_1+p_2)^2)=\langle p_1, p_2| {\cal O} |0\rangle$. As usual, the space-like
form factor $F_E(Q)=F(s=-Q^2)$ is obtained from analytic continuation.
In SCET it can be constructed from a sequence of matching calculations
with subsequent RG running, with the general result \cite{Chiu:2007yn}
\begin{equation}
 \ln F_E(Q) = C(Q) + \int_Q^M \frac{d\mu}{\mu} \left(
 \Gamma_{\rm cusp} L_Q + \gamma\right)
+ D(M) + \int_M^\mu \frac{d\mu}{\mu} \left(
 \tilde \Gamma_{\rm cusp} L_Q + \tilde \gamma \right) \,.
\end{equation}
Here $C(Q)$ is a matching coefficient at the high scale $Q$,
whose leading terms has the structure
\begin{equation}
 C(\mu)= \sum_{i=1}^3 \frac{\alpha_i(\mu)C_F^i}{4\pi} 
 \left[-L_Q^2 + \# L_Q + \# \right] + {\cal O}(\alpha_i^2) \,
\end{equation}
where $L_Q = \ln \frac{Q^2}{\mu^2}$, and 
$i=1..3$ refers to the three SM gauge group factors with
$C_F^i$ being the corresponding Casimirs. 
The numerical coefficients ($\#$) depend on the spin
of the two particles. Notice that $C(Q)$ does not depend on the
gauge-boson masses.
The RG-running between the high-energy scale $Q$ and the EW
gauge-boson mass scale $M \sim M_{W,Z}$ is controlled by the anomalous
dimension, which has a universal part, the cusp anomalous dimension
related to the Sudakov double logarithms, 
$ \Gamma_{\rm cusp}   = 4 \, \sum_{i=1}^3 \frac{\alpha_i C_F^i}{4\pi}  + {\cal O}(\alpha_i^2)$,
and a conventional part $\gamma$.

Similarly, $D(M)$ is the matching coefficient arising from integrating
out the massive gauge bosons in the SM, where the effective-theory
construction automatically takes care of the correct incorporation of
gauge-boson mixing,
 \begin{eqnarray} 
D(\mu) & = &  { \frac{\alpha_{\rm em}}{4\pi} 
 \frac{( T_3 - \sin^2\theta_W \, Q_{\rm em} )^2}{\sin^2\theta_W \, \cos^2\theta_W}}
\times 
 \left[ - L_{M_Z}^2 + 2 L_{M_Z} { L_Q} - \frac{5\pi^2}{6} \, + \, 
      \# L_{M_Z} + \# \right]
\cr 
   && \quad {} + {\frac{\alpha_{\rm em}}{4\pi} 
 \frac{T^2 - (T_3)^2}{\sin^2\theta_W}}
 \quad \times  \left[ - L_{M_W}^2 + 2 L_{M_W} {L_Q} - \frac{5\pi^2}{6} \, + \, 
      \# L_{M_W} + \# \right] + \ldots
\end{eqnarray}
A subtle point to notice is the (single-logarithmic) 
dependence of the low-energy matching
coefficient on the high-energy scale via $L_Q$, which can be traced back to the
appearance of end-point singularities in individual diagrams \cite{Chiu:2007yn}.
Finally, the RG-running in the SCET below the scale $M$ (via $\tilde \Gamma_{\rm cusp}$ 
and $\tilde\gamma$) is obtained by
replacing $\sum \alpha_i C_F^i \to \alpha_s C_F^{(3)} + \alpha_{\rm em} \, Q^2_{\rm em}$
(for QCD $\otimes$ QED).

\subsubsection{Dynamical threshold enhancement in Drell-Yan production}

\begin{figure}[t!!]
\centerline{
\includegraphics[height=0.17\textheight]{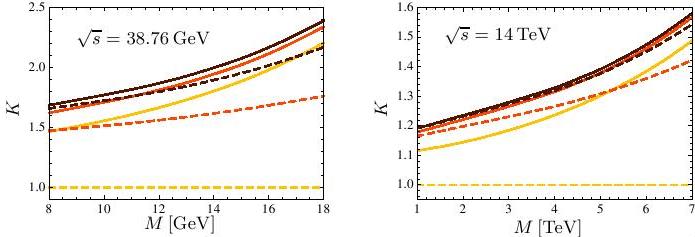}
}
\caption{\label{fig:DY} Convergence of the DY $K$-factor at 
threshold: dashed lines refer to the fixed-order
calculation (from bottom to top: LO, NLO, NNLO); solid lines
to the corresponding resummed result (taken from \cite{Becher:2007ty}).}
\end{figure}

The DY cross section in the threshold region, $z= M^2/\hat s \to 1$,
can be approximated as 
\begin{equation}
\frac{d\sigma^{\rm thr.}}{dM^2} 
 \propto \, 
 \sum_q e_q^2 \int \frac{dx_1}{x_1} \, \frac{dx_2}{x_2} 
  \, \theta[\hat s - M^2]\,
  { C(z,M;\mu_f)} 
  {\left[ f_{q/N_1}(x_1;\mu_f) \, f_{\bar q/N_2}(x_2;\mu_f) + (q \leftrightarrow \bar q) \right]}\,,
\end{equation}
where $M^2$ is the invariant mass of the DY-pair,
$x_{1,2}$ are the parton momentum fractions, and
$\hat s = x_1 x_2 s$ is the partonic c.o.m.\ energy.
For small values of $(1-z)$, we may further factorize \cite{Becher:2007ty},
\begin{equation}
  {C(z,M;\mu_f) = {} } {H(M,\mu_f)} \, {S(\sqrt{\hat s}\,(1-z);\mu_f)} \,,
\end{equation}
in order to separte the effects associated to 
the hard scale, $M^2 \sim \hat s$, set by the partonic sub-process;
the hard-collinear scale, $(1-z) M^2$, related to the virtuality
       of the colliding partons;
and a soft scale, $(1-z)^2 M^2$, related to the invariant mass of the
       hadronic remnants.
In particular, assuming a simple parametrization for the quark PDFs
at large momentum fraction,
$
 f_{q/N}(x)\big|_{x \to 1} = N_q \, (1-x)^{b_q}
$,
one can show that DY-production at threshold is dominated by $d$\/-quarks 
(which have the largest value of $b_q$), and the 
resummed $K$-factor can be written in analytic form,
from which one deduces the appearance of an effective soft scale
\cite{Becher:2007ty},
\begin{equation}
\mu_s \sim \frac{M \, (1-M^2/s)}{2 + b_d + b_{\bar d}} 
  \approx  \frac{M \, (1-M^2/s)}{13} \,.
\end{equation}
As expected, the perturbative convergence of the $K$-factor 
is significantly improved compared to the fixed order results,
see Fig.~\ref{fig:DY}.

\section{Summary}

Soft-collinear effective theory  helps:
to separate dynamical effects related to different
energy-momentum scales appearing in processes
involving soft and energetic (but low-virtuality) particles;
to establish the corresponding factorization theorems;
to define/identify process-independent non-perturbative
        input parameters/functions;
and to resum large logarithms in RG-improved perturbation theory.
Among the most important applications in inclusive
$B$-decays are the precise determination of $|V_{ub}|$ 
from $B \to X_u \ell \nu$ and SM precision tests in $B \to X_s\gamma$
(see section~\ref{sec:incl}).
 Factorization theorems in exclusive decays reduce
the non-perturbative input to universal transition form factors
and process-independent LCDAs, which can be studied by standard
non-perturbative methods (section~\ref{sec:excl}). 
Finally, SCET can be used for
systematic studies of radiative corrections in collider processes,
like EW Sudakov effects, Drell-Yan production
at threshold (section~\ref{sec:ew}), and also 
top-quark jets \cite{Fleming:2007xt},
4-Fermion Production near $WW$-threshold
\cite{Beneke:2007zg}, and traditional QCD applications 
in jet physics and parton showers \cite{Bauer:2008}.



\end{document}